\documentclass[graybox]{svmult}

\usepackage{type1cm}        
\usepackage{makeidx}         
\usepackage{graphicx}        
\usepackage{multicol}        
\usepackage[bottom]{footmisc}

\usepackage{newtxtext}       %
\usepackage{newtxmath}       

\usepackage{csquotes} 

\makeindex             

\begin{document}

\title*{Visualization Psychology for Eye Tracking Evaluation}
\titlerunning{Visualization Psychology for Eye Tracking Evaluation}
\author{Maurice Koch, Kuno Kurzhals, Michael Burch, and Daniel Weiskopf}

\institute{Maurice Koch \at University of Stuttgart, Germany, \email{Maurice.Koch@visus.uni-stuttgart.de}
\and Kuno Kurzhals \at University of Stuttgart, Germany, \email{Kuno.Kurzhals@visus.uni-stuttgart.de}
\and Michael Burch \at University of Applied Sciences Graubünden, Switzerland, \email{Michael.Burch@fhgr.ch}
\and Daniel Weiskopf \at University of Stuttgart, Germany, \email{Daniel.Weiskopf@visus.uni-stuttgart.de}
}
%
%
\maketitle

\abstract{Technical progress in hardware and software enables us to record gaze data in everyday situations and over long time spans. Among a multitude of research opportunities, this technology enables visualization researchers to catch a glimpse behind performance measures and into the perceptual and cognitive processes of people using visualization techniques. The majority of eye tracking studies performed for visualization research is limited to the analysis of gaze distributions and aggregated statistics, thus only covering a small portion of insights that can be derived from gaze data. 
We argue that incorporating theories and methodology from psychology and cognitive science will benefit the design and evaluation of eye tracking experiments for visualization.
This book chapter provides an overview of how eye tracking can be used in a variety of study designs.
Further, we discuss the potential merits of cognitive models for the evaluation of visualizations.
We exemplify these concepts on two scenarios, each focusing on a different eye tracking study.
Lastly, we identify several call for actions.}

\newpage
\section{Introduction}
\label{sec:wws_introduction}

\begin{figure}[t]
    \centering
    \includegraphics[width=\textwidth]{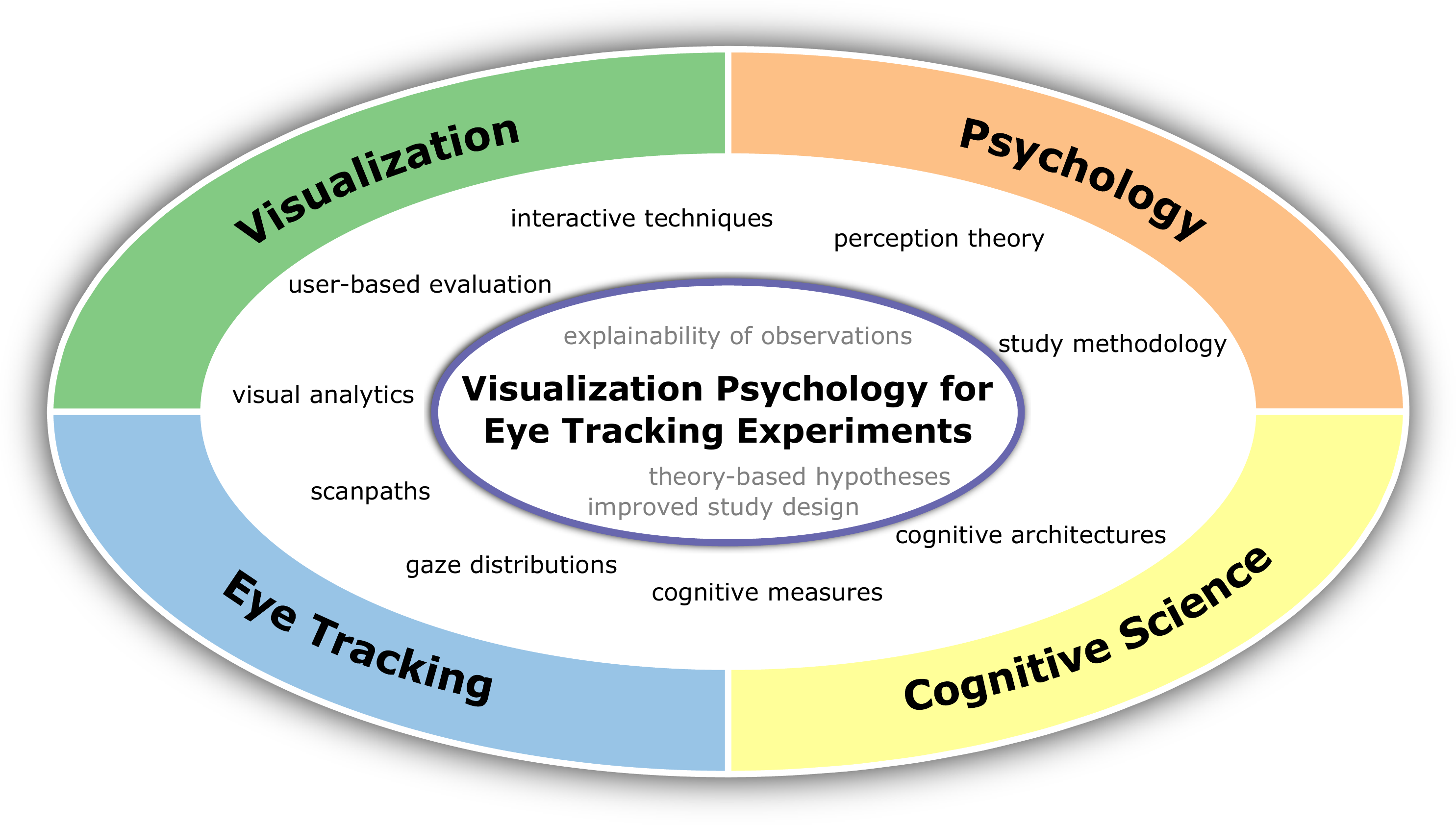}
    \caption{Visualization psychology for eye tracking experiments incorporates expertise from psychology and cognitive science to improve the evaluation of visualization techniques by study methodology, theory integration, and cognitive architectures. Figure reprinted from Kurzhals et al.~\cite{kurzhals2020beyond}.}
    \label{fig:teaser}
\end{figure}

Eye tracking experiments in visualization research provide insights into how people interpret and interact with visualizations. 
In contrast to classic performance analysis, the analysis of gaze behavior provides information about the distribution of visual attention over time. Eye tracking further helps understand visual strategies employed in interpreting a visualization or in working with a complex visual analytics system. 
In addition, machine learning, statistics, visualization research, and data science in general contributed a multitude of new techniques~\cite{blascheck2017, Duchowski2017} to expand the spatio-temporal analysis of eye tracking data, verify results, and formulate new hypotheses.
By combining such state-of-the-art analysis techniques with expertise from psychology, cognitive science, and eye tracking research, as depicted in Figure~\ref{fig:teaser}, the design and insights gained from eye tracking experiments in visualization can be significantly improved.
However, evaluation in visualization still lacks concrete guidance on such interdisciplinary research.
One part of the problem is the increasing disconnect between psychology and visualization research.
For example, in visual analytics, there is less focus on individual visualizations but on the processes that the tool is meant to support.
Such processes often can be related to different scenarios, such as visual data analysis and reasoning and collaborative data analysis \cite{Lam2011}, to name a few.
Although visualization research has become more process-centered on a conceptual level, evaluation today still mostly involves usability testing and benchmarking based on completion time and error metrics.
For this reason, we advocate that the visualization community broadens their scope toward evaluation methodologies that better capture the dynamics of complex tool interactions.
In a similar sense, we advocate that cognitive psychologists actively participates in that endeavour by focusing their study on higher-level cognition.
Fisher et al. \cite{Fisher2011} even call for translational research that bridges pure science and design, with the hope to better support knowledge transfer between both fields.
A major inspiration for this work has been Kurzhals et al.~\cite{kurzhals2020beyond}, who advocated for more interdisciplinary research between the fields of psychology, cognitive science and visualization.
In this book chapter, we exemplify how the eye tracking modality could be beneficial to a broader scope of empirical studies, beyond classical laboratory experiments.

\section{Study Designs}
\label{sec:study_designs}

In the following, we describe how different study designs commonly found in visualization evaluation~\cite{Carpendale2008} can benefit from eye tracking methodology.
Eye tracking has become popular in the evaluation of visualizations and there is wide variety of methods and metrics to evaluate the performance of visualization~\cite{Goldberg:11}.
Kurzhals et al.~\cite{Kurzhals2016} reviewed 368 publications that include eye tracking in a user study and identified three main approaches to evaluate visualizations: evaluating the distribution of visual attention, evaluating sequential characteristics of eye movements, and comparing the viewing behavior of different participant groups.
Their review also shows that user studies with eye tracking have become more common in recent years.

However, the use of eye tracking in evaluation methods has been narrow in the sense that it is predominantly used in laboratory experiments but infrequently found in in-the-wild studies.
Laboratory experiments offer great control and precise results, but are primarily suited to study individual factors with predefined hypotheses.
In this section, we outline the current practice of using eye tracking in visualization research, mostly in the context of controlled experiments.
Furthermore, we outline how eye tracking could be beneficial beyond laboratory experiments.
For this, we include a discussion of in-the-wild studies.

\subsection{Controlled Experiments}

Eye tracking has become increasingly popular in laboratory experiments.
In visualization research, controlled experiments have been mostly conducted for summative evaluation, such as usability testing and benchmarking.
However, such studies often fail to relate their findings to the underlying cognitive processes.

Here, we showcase just a few selected eye tracking studies in visualization with a strong focus on cognitive aspects, such as reasoning, memorability, and perception.

Huang et al.~\cite{Huang2009} studied how link crossings in graph drawings affect task performance.
Participants were asked to find the shortest path between two specified nodes for each drawing.
Their eye tracking experiment revealed that link crossings, contrary to the common belief, only have minor impact on graph reading performance, especially at angles of nearly 90 degrees.
Instead, the extra time spent on certain drawings was due to the tendency of subjects to prefer certain paths at the beginning of the search task.
It was observed that subjects tend to follow links that are close to an (imaginary) straight line between the target nodes.
This can increase the search time if no such links exist in the graph drawing, and alternative graph lines must be considered.
This behavioral bias during the initial search process in graph drawings was termed geodesic-path tendency.
Körner et al.~\cite{Koerner2011, Koerner2014} found that this behavior can be explained by studying to which extent search and reasoning processes in graph comprehension are performed concurrently.
The two main process involved in such a task are first detecting both specified nodes in the graph (search) and next finding the shortest path between those two nodes (reasoning).
Assuming that these processes occur in parallel, subjects would not show this kind of bias toward certain links in graph drawings as described by geodesic-path tendency.
Körner et al. conducted eye tracking experiments and found that these two graph comprehension processes indeed are mostly performed sequentially.
This means that subjects can only rely on local information of the graph drawing to perform reasoning during the search task.

Borkin et al.~\cite{Borkin2016} studied the memorability of visualizations and how well they are recognized and recalled.
Their experiments consists of three phases: encoding, recognition, and recall.
In the encoding phase, subjects were exposed to 100 different visualizations sampled from the MassVis dataset.
After the encoding phase of 10 seconds per image, subjects were exposed to the same images plus unseen filler images as part of the recognition phase.
In both phases, eye fixations were collected to examine the elements in visualizations that facilitate the memorability.
In the last phase, subjects were asked to describe correctly identified images as best as possible to understand what elements were easily recalled from memory.
In the encoding and recognition phases, eye fixations were analyzed with heatmaps to find out what parts of the visualization draw initial attention to subjects.
During encoding, subjects tend to perform visual exploration, and fixations are distributed across the image.
This pattern can be observed on most images. 
Fixations during the recognition phase are distinct between most recognizable images and least recognizable images.
It was shown that in the most recognizable visualizations, fixations are more biased toward the center of the image and are generally less widely distributed.
This means that relatively few fixations are needed to recall easily recognizable images from memory, whereas less recognizable images require more contextual information.
Their study also shows that participant descriptions are of higher quality for visualizations that are easily recognizable even with a reduced amount of encoding time (such as one second).
Interestingly, prolonged exposure does not change the fact that some visualizations stay more recognizable.

Hegarty et al.~\cite{Hegarty2010} studied how saliency of task-relevant and task-irrelevant information on weather maps impacts task performance.
Mean proportion of fixation time was measured to study the level of attention on task-relevant or task-irrelevant information before instructions and after-instructions.
On the one hand, it was reported that fixation time significantly increases on task-relevant areas after instructions were given, which shows that attention is strongly driven by top-down influences. 
On the other hand, visual salient regions do not draw attention to participants, unless they correspond to task-relevant areas.
These results emphasize that visual salience does not necessarily facilitate task performance, unless participants are sufficiently guided by top-down processes toward task-relevant information.

The aforementioned visualization studies exemplify that eye tracking has become an established modality to study cognitive processes.
Furthermore, many of these results are directly applicable to the visualization community 

\subsection{In-the-Wild Studies}
\label{subsec:inthewild}

\begin{figure}[t]
    \centering
    \includegraphics[width=0.8\textwidth]{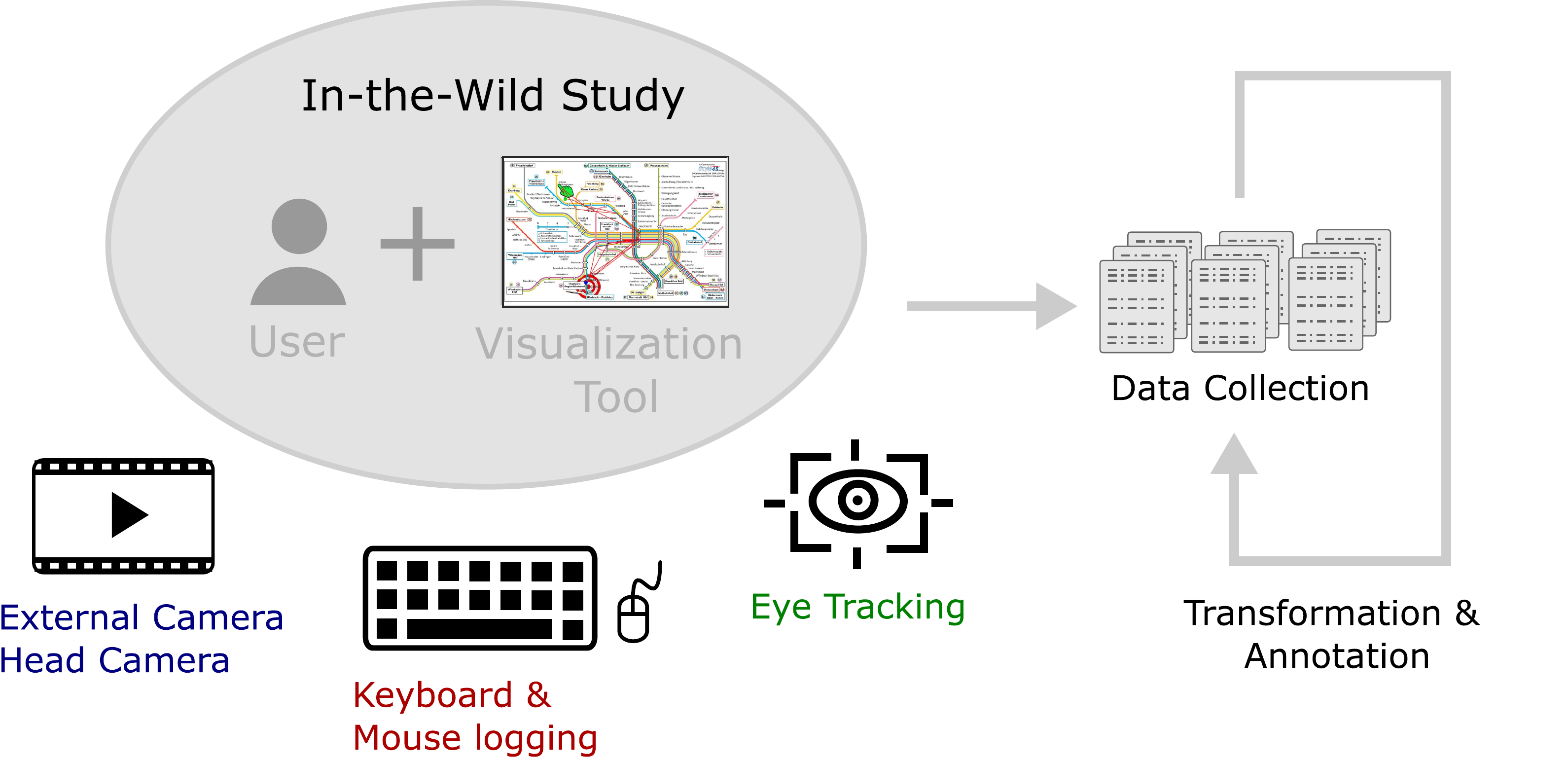}
    \caption{Conceptual overview of data collection in in-the-wild studies. Different modalities such as camera views, keyboard and mouse logging and eye tracking can be combined to generate a data-rich description of the study. The collected data can be transformed or extended semantically with labels provided by human annotators. The small visualization inset shows a figure reprinted by permission of Taylor \& Francis Ltd from Netzel et al.~\cite{Netzel_SCC:17}.}
    \label{fig:overview-in-the-wild}
\end{figure}

As the complexity of visual artefacts increases, it becomes harder to provide holistic assessments of the effectiveness of complex visualization tools.
Field studies offer more realism by assessing systems within their natural environment like at the domain expert's work place.
In such settings, it is easier to study processes, like sense-making, since they tend to be highly context-sensitive.
Thus, such processes are more difficult to capture in controlled experiments that usually impose tight protocols~\cite{Lam2011}.
Many researchers believe that visualization evaluations could benefit from more qualitative research, for example, by employing ethnographic techniques~\cite{Shneiderman2006, Ellis2006}. In general, social science methods should receive more attention in the community since individual assessment techniques often fail to capture contextual factors~\cite{mcnamara2011workshop}.

Ethnographic techniques have been advocated by Sheiderman et al.~\cite{Shneiderman2006} in the form of multi-dimensional in-depth long-term case studies (MILCs).
MILCs are performed in-field, in a domain experts natural working environment, thus they are unobtrusive and guarantee more realistic results.
Data collected in MILCs is mostly qualitative and consists of interviews, log books, user maintained diaries, and usage statistics obtained from the visualization tool.
Field studies are often based on ethnographical participant observation methods, interviews, surveys, and automated logging of user activity~\cite{Shneiderman2006}, i.e., they are predominantly qualitative research in terms of data collection and analysis.
Qualitative evaluation often involves thematic analysis and manual coding, both are inherently subjective processes~\cite{Carpendale2008}.
There are multiple problems associated with a primarily quantitative data collection and analysis approach.
First, data collection and analysis are tedious processes that often involve a lot of manual work.
In terms of data analysis, software tools like computer-assisted qualitative data analysis software (CAQDAS) \cite{Blascheck2016} improve the efficiency of thematic analyses and assist coding, but only to a limited extent.
This problem gets exacerbated in long-term studies where a large amount of diverse data is collected.
For this reason, many MILCs come only with a few interviews and observations, and during the study, data collection is sparse, at most it consists of user interface logs that are automatically recorded (in practice, even logging is very uncommon except for Sheiderman's MILC study~\cite{Shneiderman2006}).

The usage of physiological sensors is in particular challenging in ethnographic studies, where the property of unobtrusiveness must be obeyed (interference by study coordinators needs to be kept minimal).
This is hardly achievable with stand-alone eye tracking devices and electroencephalogram (EEG), which are highly invasive and lack mobility.
Furthermore, such physiological sensors often require external supervision and careful setup.
This naturally restricts what and how data is collected in ethnographic studies.
However, in regard of eye tracking devices, we have seen technological progress toward mobile devices that are less invasive and require almost no external supervision.
In this way, eye tracking could act as a quantitative modality that does not interfere with ethnographic requirements like unobtrusiveness.
Figure \ref{fig:overview-in-the-wild} illustrates the basic idea of collecting data from multiple sources and semantically and/or algorithmically extending it in subsequent steps.

Whether a modality is considered invasive depends not only on the modality itself, but also on the situational context.
For example, think-aloud protocols can be elicited either naturally, or they can be imposed externally on request (by a study coordinator), which could negatively affect reasoning processes~\cite{Hernandez2011}.
Think-aloud might also negatively interfere with the natural eye movement, for example, during attending the screen.
To compensate this issue retrospective think-aloud \cite{Kurzhals2016} of screen recordings accompanied by eye tracking data was suggested \cite{Elling2011}.
In general, it is important to detect these attention shifts, which also occur naturally without external stimulation and revalidate the recorded eye movements.
Transferring our studies to virtual reality (VR) could provide non-invasive access to physiological sensors that are readily available in VR headsets.
This could go beyond eye tracking and further include tracking head/body movements and interface interactions.

The previously discussed scope of in-the-wild studies is on individuals, but can be easily extended to collaborative settings as well.
In that regard, pair analytics~\cite{Hernandez2011} provides an interesting approach to studying social and cognitive processes for the evaluation of visual analytics tools.
Pair analytics studies the interaction between two human subjects, the subject matter expert and the visual analytics expert, and visual analytics tools.
The visual analytics expert and  subject matter expert collaborate to solve a specific domain goal, but both have different responsibilities and roles in that process.
The  subject matter expert (driver role) is the domain expert that has the contextual knowledge but not the expertise to use the visual analytics tools, whereas the visual analytics experts (navigator role) lacks the domain knowledge but the technical expertise to translate the verbal requests from the subject matter expert to tool commands.
The dialog between the subject matter expert and visual analytics expert makes the mental models and cognitive processes explicit, thus captures important cues of the collaborative process.
Compared to classical think-aloud protocols, verbalization during collaborative processes occurs naturally.
Aligning the rich data from think-aloud protocols with eye-movements from the subject matter expert and visual analytics expert could be a good starting point for in-depth analysis on social and cognitive processes.
Kumar et al.~\cite{Ayush2020} have proposed a similar type of study, but in the context of pair programming.
Data from eye tracking data and other modalities, like recorded video, are time-synchronized.
Having discussed the merits of in-the-wild studies in the evaluation of visualizations, we also need to the address the inherent difficulties of conducting those studies.
As Shneiderman et al.~\cite{Shneiderman2006} already mentioned, it is necessary for researchers and participants to allocate a considerable amount of time into such studies.
For example, Valiati et al.~\cite{Valiati2008} performed multiple longitudinal case studies, each took about three to four months.
This complicates recruiting participants, in particular, when domain experts are needed.
It needs to emphasized that this requires an intense level of collaboration and devotion from both the researchers and domain experts.

\subsection{Bridging between Quantitative and Qualitative Research}

The aforementioned study designs can be roughly classified as being either qualitative or quantitative.
Quantitative evaluation, often in laboratory experiments, follows statistical frameworks to make precise inferences about predefined hypotheses.
Qualitative evaluation provides a richer understanding of the situation that is more holistic than what quantitative evaluation can capture, but also less precise \cite{Carpendale2008}.

Study designs that encompass data collection, analysis, and inferences techniques from both methodological paradigms, can potentially offset their individual shortcomings.
The commonly found dichotomy in quantitative and qualitative inquiry is too narrow.
This motivates the research field of mixed methods, which uses methods from both disciplines to provide a better understanding of the studied phenomena \cite{Johnson2007}.
One of the hallmarks of mixed methods is to achieve integration by bringing qualitative and quantitative data together in one study \cite{Moseholm2017}.
This integration can occur at different levels such as integration at the study design level, methods, and interpretation/reporting. 
An example of integration at study level is an explanatory sequential design where the quantitative phase informs the follow-up qualitative phase.
For example, a controlled study design with eye tracking could be conducted to quantitatively evaluate the performance on a visual search tasks with two different visual representations. A follow-up qualitative phase could be justified for several reasons. For example, a group of participants could strongly deviate in performance. The follow-up qualitative phase could try to identify the root of this cause by performing a retrospect think-aloud protocol where the respective participants comment on their played-back eye-movements.
Think-aloud can also be performed concurrently to eye tracking experiments, which would correspond to a convergent mixed methods design.

Integration at the other two levels is more concerned with mixed data analysis and it is considerably more challenging and less explored \cite{Moseholm2017, Vogl2019}.
Common strategies of mixed-data analysis include: data transformation, typology development, extreme case analysis, and data consolidation \cite{Caracelli1993}. 
Data consolidation is one of the greatest challenges of mixed-data analysis since it merges two data sets, which goes beyond linking.
The difference is that both data sources remain clearly identifiable after data linking while consolidation leads to a genuine new piece of information.
These techniques are not necessarily distinct, for example data transformation could be an important prepossessing step for data consolation.
Data transformation encompass two data conversion directions, either quantified data is transformed to qualitative data (qualtizing) or vice versa (quantizing) \cite{Vogl2019}.
A common way to perform quantization is by counting codes in an thematic analysis. 
In that way, quantitative methods like inferential statistics can be applied indirectly to qualitative data.
Qualtizing can be seen as a semantic transformation of the original quantitative data.
This could add a semantic link to quantitative measurements, which is usually not present in such measurements beforehand.
For example, gaze data in its raw form is just a trajectory in 2D space without any semantic link to the underlying stimulus.
For static stimuli, this semantic link is easy to provide since there is a one-to-one correspondence between gaze location and stimuli location.
However, such a direct correspondence it not present in dynamic stimuli where the underlying scene is varying over time.
Providing additional semantics to gaze data with underlying dynamic stimuli, for example, by labeling time spans according to the participant's activity, would increase the usefulness of these measurements.
This form of data consolidation by annotation of quantitative data can improve the credibility of those measurements and thereby improve the quality of subsequent mixed data analysis steps.

\section{Explainability of Observations} 

As already outlined in the previous section, building semantic links between gaze data and contextual factors, like scene information or activity labels, can aid the data analysis and thereby the explainability of observations.

\paragraph{Areas of Interest}
Scanpaths can be transformed to qualitative data by mapping each fixation to a label, which uniquely identifies an area of interests (AOIs).
The usefulness of such a representation depends on the semantics of AOIs.
For example, AOI grids automatically generated for static stimuli do not provide much semantic details since an AOI hit is just still just an indicator of spatial position (spatial quantization), but does not provide semantic information w.r.t the underlying visual entity.
A similar problem occurs for AOIs induced by automatic clustering of gaze data, where regions with strong accumulation of gaze positions are defined as AOIs.
In contrast to such automatically generated AOIs, manually AOIs defined based on semantics (images on web pages; axes on graphs; etc.) can provide more detailed information.

\paragraph{Interpretation and Data Analysis}

In Section~\ref{sec:study_designs}, we have mentioned the challenges in data collection and analysis in the context mixed-methods research.
These kind of challenges are particularly relevant for in-the-wild studies, such as the previously described long-term field studies in pair analytics.
It is challenging to integrate data from heterogeneous data sources, such as eye tracking and other physiological sensors, as well as hand-written or verbal protocols.
An interesting approach toward these problems is visual data analysis, sometimes referred to as \textit{visualization for visualization (Vis4Vis)} \cite{Weiskopf2020}.
The vision behind Vis4Vis is to use visualizations to analyze and communicate data from empirical studies.
In the context of eye tracking studies, visual analysis tools have shown to support the evaluation of studies.
For example, Blascheck et al.~\cite{blascheck2014state} provide a comprehensive overview of visualization techniques for eye tracking data.
Some visual analysis approaches have been proposed that integrate eye tracking data with other data modalities, such as think aloud protocols and interaction logs.
Blascheck et al.~\cite{Blascheck2016} proposed a visual analytics system that allow interactive coding and visual analysis of user activities.
Such approaches could be considered as a first step toward visual analysis of data-rich empirical studies with multiple data modalities.
Nonetheless, there is still the need for more scalable visual representations and automatic analysis techniques to better support the analysis of data from long-term empirical studies.

\section{Cognitive Architectures} 

\begin{figure}[t]
    \centering
    \includegraphics[width=0.8\textwidth]{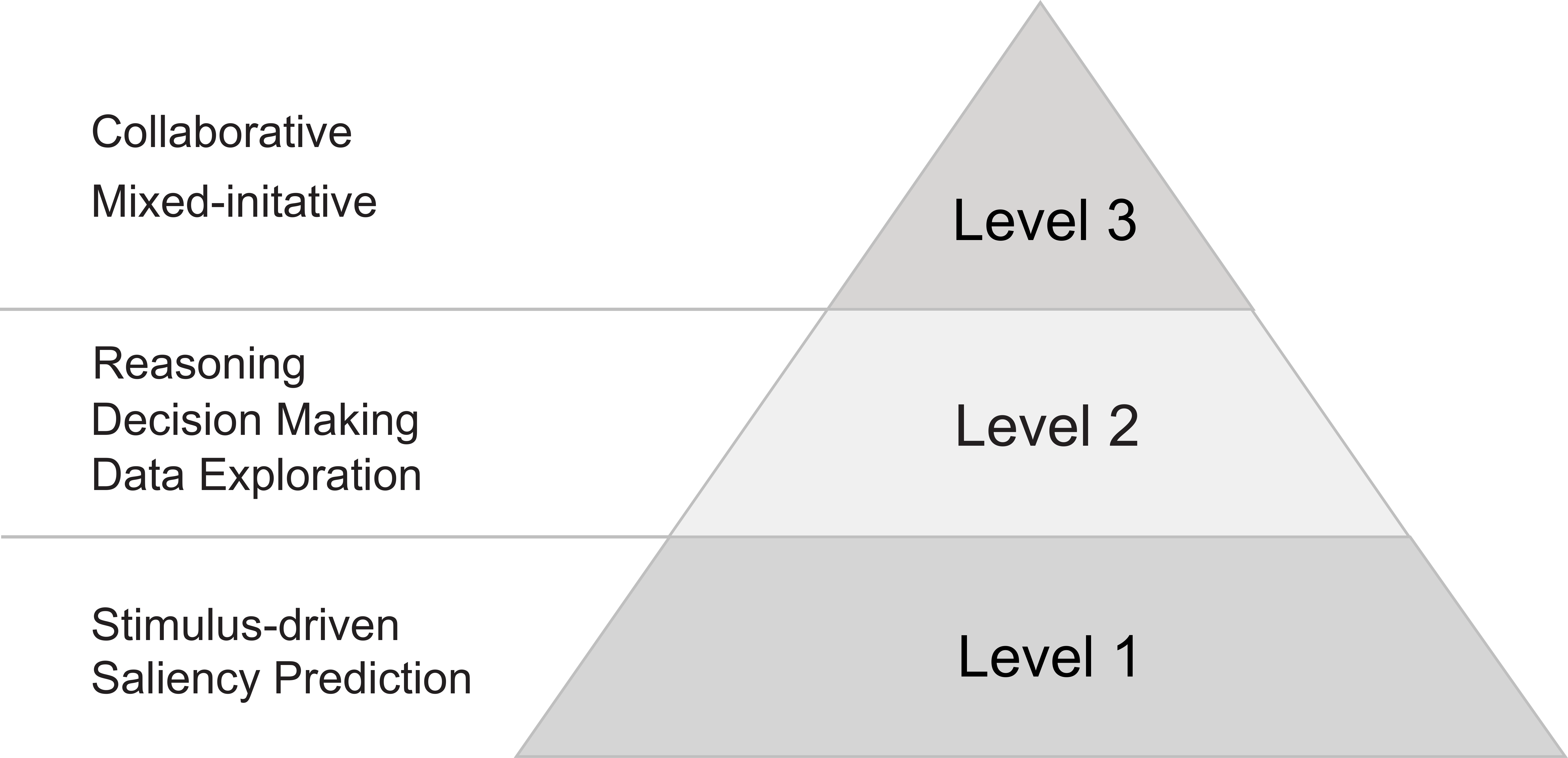}
    \caption{Cognitive simulation can be performed on multiple levels. Each layer corresponds to one class of tasks. Each level depends on its lower levels. For example, simulation of collaborative settings with multiple individuals performing a common task requires successful simulation of cognitive tasks (level 1 and 2) for individuals.}
    \label{fig:cogn_arch_pyramid}
\end{figure}

One of the overarching goals of empirical studies in visualization is to formulate guidelines and heuristics that inform the design of future visualizations.
However, many psychological phenomena only apply to specific aspects of the evaluation, like Gestalt Laws, but visualization consists of multiple perceptual and cognitive aspects combined.
Thus, guidelines and heuristics on system level would be preferable.
However, since they typically involve higher-level cognitive tasks, they are more influenced by individual factors, such as knowledge, cultural background, and cognitive capabilities.
Computational models have the potential to generalize across a wide range of individuals \cite{Kurzhals2016} and can provide methods to accurately predict the effectiveness of visual designs \cite{Heine2020}.
As shown in Figure \ref{fig:cogn_arch_pyramid}, such simulation could be performed on multiple levels.
On the most fundamental level one, simulation of human cognition boils down to perceptual simulation that is often highly driven by the stimulus or more general bottom-up influences.
Early work on that level has been proposed by Itti and Koch in the context of visual saliency prediction \cite{Itti2001}.
In general, cognitive simulation on higher levels has been less explored, mostly due to its complexity and the lack of formal descriptions.
Nonetheless, computational models based on cognitive architectures have been proposed to automate the evaluation of visualizations on the level of reasoning and decision making.
One example of the application of cognitive architectures like ACT-R \cite{anderson1997} is \emph{CogTool} (see https://www.cogtool.org), which is deployed for the initial validation of web designs.
Eye fixations can play an important role as a means to train and validate cognitive models.
For example, Raschke et al. \cite{Raschke2014} propose a cognitive model based on ACT-R that simulates visual search strategies. Their motivation is to build a simulation tool similar to \emph{CogTool} that allows automatic, thus non-empirical, evaluation of visualizations. In contrast to \emph{CogTool}, that is based on an extended version of Keystroke-Level-Model \cite{Card1980}, their model is trained on eye fixations. 
Although their work does not provide any concrete implementation, other researchers have demonstrated that models based on ACT-R can simulate eye movements on simple line charts with high confidence~\cite{peebles2012cognitive}.
Their model even provides vocal output, thus, is able to simulate graph comprehension with results close to human level.
From a technical viewpoint, cognitive architectures like ACT-R have some limitations that prevent their adoption to more complex tasks.
For example, Heine et al.~\cite{Heine2020} advocate the use of probabilistic models, like Dynamic Bayesian networks, in the context of modeling human cognition.
Probabilistic models could provide a unified mathematical model toward human cognition and  allows to describe variation of factors that are not explicitly modeled.
This is a strong advantage over ACT-R that depends on explicit rule-based modeling, which does not scale well for sophisticated visualizations.

\section{Example Scenarios}
\label{Examples:Sec}

Visualization evaluation could benefit from the aforementioned study designs, the explainability of observations, and cognitive architectures. We exemplify this, based on two previous eye tracking studies. One on the design of metro maps~\cite{Netzel_SCC:17} and one on the evaluation of parallel coordinates plots~\cite{Netzel:17}.
We discuss how these studies could be enhanced and extended by adopting ideas from the previous sections of this chapter.

\subsection{Overview of Scenarios}

\begin{figure}[t]
    \centering
    \includegraphics[width=\textwidth]{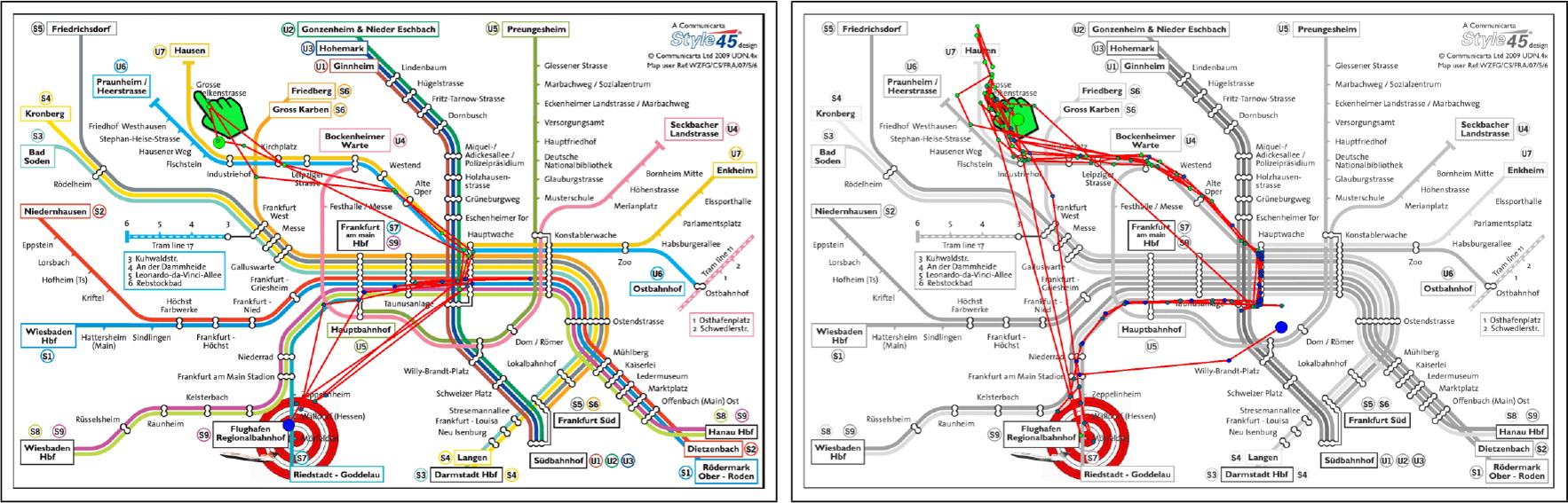}
    \caption{Scenario 1: Metro maps in color (left) and in gray scale (right) have been compared for solving a way finding task from a start (hand) to a target location. Eye tracking was measured to identify differences in the reading behavior of both conditions. Figure reprinted by permission of Taylor \& Francis Ltd from Netzel et al.~\cite{Netzel_SCC:17}.}
    \label{fig:scenario1}
\end{figure}

\paragraph{Scenario 1: Metro Maps}
Investigating the readability of metro maps is a challenging field of research, but the gained insights are valuable information on how to find design flaws, enhance the design, and make the maps more understandable to travelers~\cite{Burch_IVAPP:16}. Netzel et al.~\cite{Netzel_SCC:17} compare color-coded and gray-scale public transport maps with an eye tracking study. The major outcome is that color is an important ingredient to reduce the cognitive burden to follow lines. 
Eye tracking was essential in this study to understand the strategies participants applied to solve a route finding task between a start and a target station (Figure~\ref{fig:scenario1}).
The analysis showed that color maps led to much longer saccades, and it was hypothesize that colored lines made participants feel safe and, hence, the route finding tasks could be answered faster and more reliably. In contrast, in gray-scale maps, the participants' eyes moved with significantly smaller saccades to trace a line reliably, which was due to missing color that would otherwise have helped to visually and perceptually separate the metro lines from each other. A practical result of this eye tracking experiment for the professional map designer is that color is crucial for route finding tasks, hence the much cheaper printed variants in gray-scale would obviously be counter-productive for the business, although the costs are much lower.

\begin{figure}[t]
    \centering
    \includegraphics[width=\textwidth]{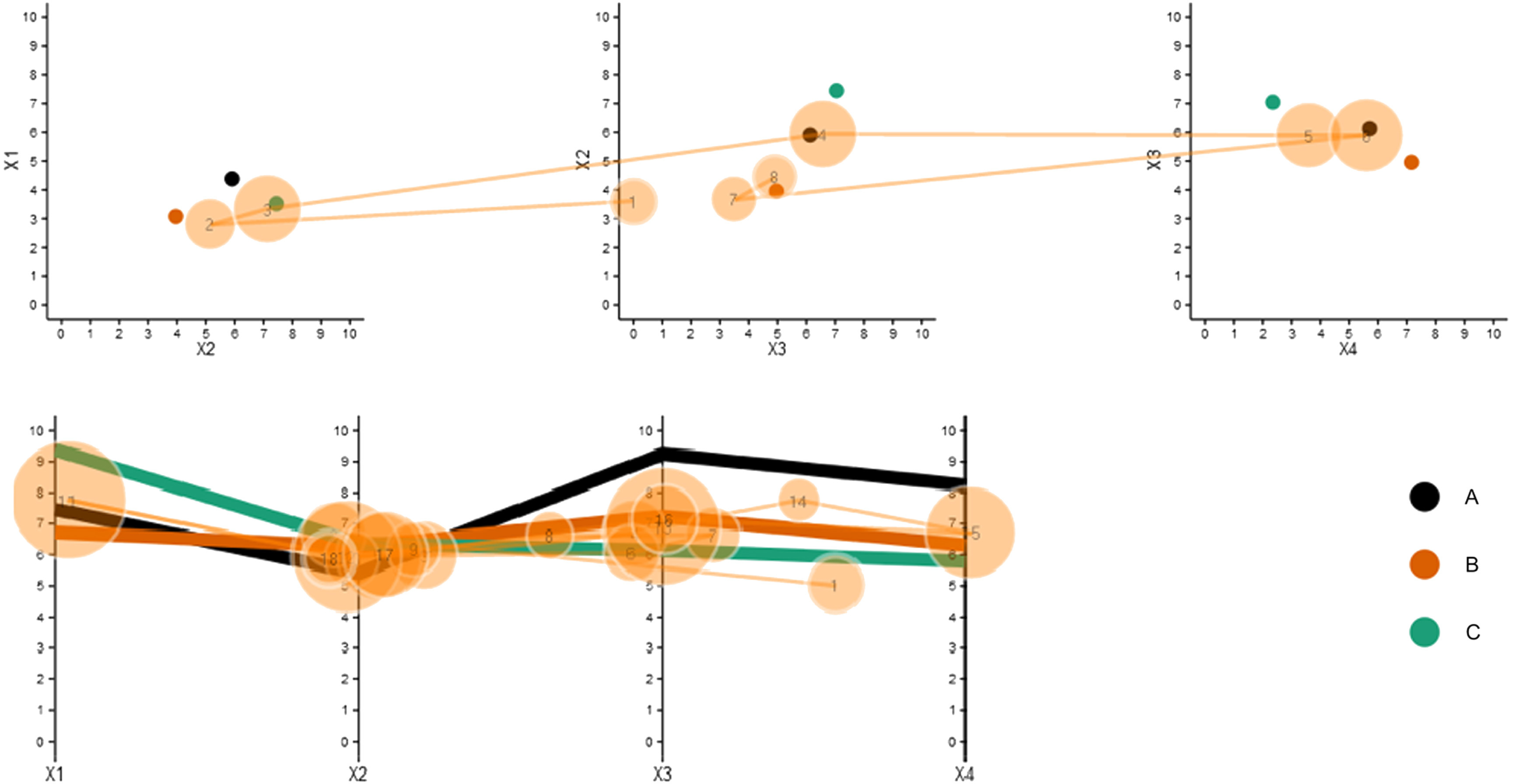}
    \caption{Scenario 2: Eye tracking was applied to compare how people compare distances between three points (A, B, C) in scatterplots (top) and parallel coordinates plots (bottom). Figure reprinted by permission of Elsevier from Netzel et al.~\cite{Netzel:17}. Copyright 2017 Zhejiang University and Zhejiang University Press. Licensed under the CC BY-NC-ND 4.0 license (https://creativecommons.org/licenses/by-nc-nd/4.0/).}
    \label{fig:scenario2}
\end{figure}

\paragraph{Scenario 2: Scatter and Parallel Coordinates Plots}
The second example of a study investigates the assessment of relative distances between multi-dimensional data points with scatterplots and parallel coordinates plots~\cite{Netzel:17} (Figure~\ref{fig:scenario2}). The authors performed an eye tracking study and showed that scatterplots are efficient for the interpretation of distances in two dimensions, but participants performed significantly better with parallel coordinates when the number of dimensions was increased up to eight. With the inclusion of eye tracking, it was possible to identify differences in the viewing of the two visualization types considering fixation durations and saccade lengths.
The authors further introduced a visual scanning model to describe different strategies for solving the task. With the help of eye tracking, a bias toward the center (parallel coordinates plot) and the left side (scatterplots) of the visualizations could also be measured, which is important for the design of such plots considering where participants will potentially spend most of their attention. However, understanding clear visual attention patterns like following a line as described in the former eye tracking study is not possible here since either the diagram consists of crowds of points (scatterplot) or a lot of crossing and partially occluding polylines (parallel coordinates plot). Hence, the reading behavior is more complex and harder to model than in Scenario 1.

\subsection{Potential Extensions}

\paragraph{In-the-Wild Studies}
As described in Section~\ref{subsec:inthewild}, studies in the wild provide a higher realism for experimental outcomes. For Scenario 1, this is highly desirable because the interpretation of metro maps is a task performed by many people in everyday situations. For the sake of controlability, stimuli and task were adjusted to fit to a laboratory setting: People were watching metro maps on a screen with start and goal clearly highlighted. The situation in a real metro station would differ significantly. Numerous confounding factors such as distractions by other people, no clear identification of start and goal, as well as other potential stress inducing factors might influence the results how people look at such a map.

Scenario 2, in contrast, involves visualization techniques (i.e., parallel coordinates plots) that are less known to people. An application in the wild would presumably take place with domain experts and data scientists rather than a more general audience of students, as it was the case in the conducted study. Further, the set of performed tasks would be extended in comparison to the lab study.
However, for the hypotheses of the original experiment, the expertise of the participants was not the determining factor, since the study aimed to analyze general behavior. For measurements over longer time periods, the experts could potentially show additional behavior patterns and learning effects, while general behavior aspects should not change.

\paragraph{Collaborative Studies and Pair Analytics}
The investigation of metro maps in Scenario 1 is often an individual task, but is in real life also performed collaboratively. Similar to the application of the task in the wild, the analysis of collaborative task solving has the potential to reveal details on how decision making is performed.
Scenario 2 can be imagined for typical analysis tasks involving domain and visualization experts.
In both scenarios, the dialog between participating people provides valuable information on a qualitative level. 
Scenario 1 provides the possibility to perform a symmetrical setup where both persons have the same prerequisites and solve the task together.
In Scenario 2, the integration of the visualizations in a visual analytics framework has the potential to focus more on a pair analytics approach where people with different fields of expertise (i.e., domain and visualization expert) work together to solve the task.

Further, measuring the gaze behavior of both persons indicates periods when they potentially share visual attention, and when they might be confused, e.g., searching for the region the other person is talking about. Hence, eye tracking helps evaluating the visualization at hand, but also the interaction between persons.

\paragraph{Mixed Methods}
Qualitative and quantitative evaluation combined provide a more comprehensive understanding of the research topic than each method on its own. Scenario 1 and 2 mainly focused on the quantitative evaluation of traditional performance measures and established eye tracking metrics. However, with respect to the analysis of visual strategies, both studies included visual analysis for the qualitative assessment of recorded scanpaths. We argue that such observations will become more important for experiments whenever eye tracking is involved. Furthermore, additional data (e.g., think aloud, interaction logs) will be necessary to include in a data integration step to provide a new, more thorough view on the participant's behavior. 

\paragraph{Cognitive Models}
Cognitive models to predict the scanpath of a participant and the efficiency of wayfinding tasks would be beneficial for the design of metro maps in Scenario 1. Although different strategies for solving the task could be identified, a generalized model was not included in the results of the study. The study was one of the first in this domain where it was important to identify general strategies. For a comprehensive model, additional data for different levels of expertise might be necessary. Here, map designers and map readers are two different target groups that potentially focus on different aspects of the map and viewing tasks might differ significantly between such groups.
An implicit model of strategies was applied for the manual annotation of paths, imprecise measures of line tracing. Future models could also consider psycho-physical measures, for example, just noticeable differences to be able to separate close-by metro lines. In the wild, saliency models will also play an important role for the orientation while searching for start and goal locations.

\begin{figure}[t]
    \centering
    \includegraphics[width=0.8\textwidth]{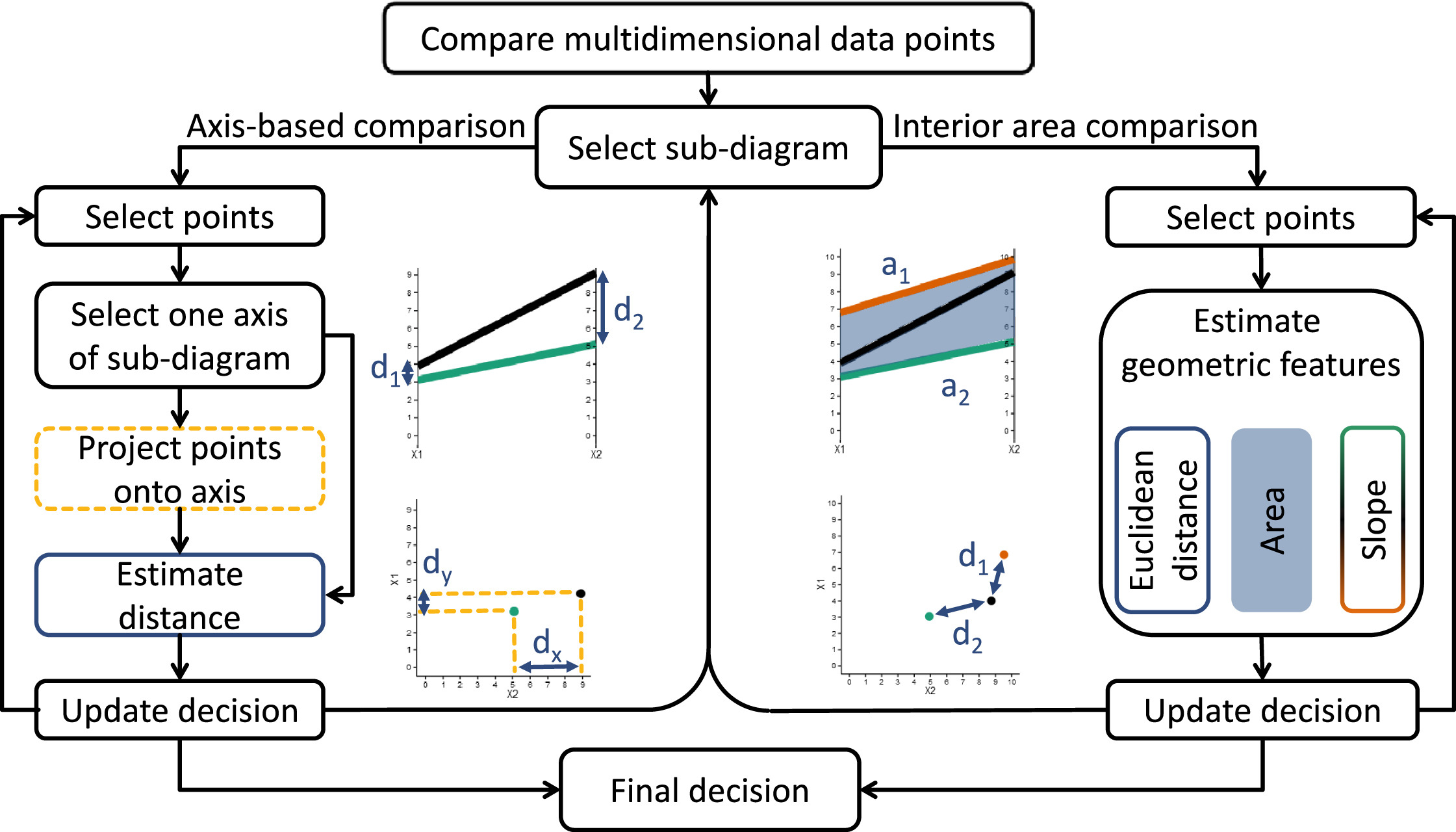}
    \caption{Strategy model for the visual comparison of multidimensional data points with parallel coordinate plots. Netzel et al.~\cite{Netzel:17} identified two strategies, i.e., axis-based and interior area comparison, and comprised them in a hand-crafted behavior model. Figure reprinted by permission of Elsevier from Netzel et al.~\cite{Netzel:17}. Copyright 2017 Zhejiang University and Zhejiang University Press. Licensed under the CC BY-NC-ND 4.0 license (https://creativecommons.org/licenses/by-nc-nd/4.0/).}
    \label{fig:model}
\end{figure}

The design of the study in Scenario 2 was based on some assumptions made from theory and observations in pilot experiments. Netzel et al. provided a handcrafted model (Figure~\ref{fig:model}) on the different strategies during the reading process of the visualization. This model was guided by the hypotheses of the study.
In future research, such models could be generated more systematically, informed by theoretical perceptual or cognitive models from psychology.

\newpage
\section{Call for Actions}

Based on our previous observations, we have identified the following interesting points for future development and calls for actions. 

\paragraph{Translational research}

Many early guidelines in visualization were informed by perceptual and cognitive science, like efficient visual encoding, Gestalt laws~\cite{Koffka, Wertheimer}, or feature integration theory~\cite{Healey:12, Treisman:85}.
However, there is lack of guidelines that inform design decisions for visual analytics systems \cite{Scholtz2014}, since current cognitive models are good at explaining cognitive processes on well-defined tasks and simple visual stimuli, but are less applicable to the aforementioned scenarios that have become prevalent in today's systems~\cite{Hegarty2011}.
This line of research offers great potential for translational studies since psychology and visualization research would equally benefit from such results.
Distributed cognition could be a promising approach toward translational studies of that kind since it provides a more holistic view of the way humans reason and think.
It acknowledges the fact that humans live in materialistic and social environments, thus, it emphasizes the importance of contextual factors in human cognition~\cite{Hutchins2000}.

\paragraph{Best Practices}

This book chapter only provided a high-level conceptual view on evaluation strategies.
So far, our envisioned evaluation strategies have not yet been implemented in real world empirical studies.
Many challenges are left unanswered, as how to practically design, conduct, and evaluate data-rich empirical studies.
It is particularly important to provide researchers a tool set to perform sophisticated data analysis with minimal effort.
There is also need for the whole community of researchers to agree upon a proper way to report results of such studies.

\paragraph{Interdisciplinary Research Venues}

Psychologists' core topics are often disconnected from topics relevant for visualization research. Yet, there are some successful examples of combining communities,  for example, at the \textit{Symposium on Eye Tracking Research and Applications (ETRA)}. Such events provide great opportunities for interdisciplinary discourse and establishing collaborations. However, publication strategies and research topics might significantly differ between communities.
Hence, a fusion of expertise just by project collaborations might cover some research questions, but from a long-term perspective, other solutions are necessary.
A key question, of course, is: \textit{How can we integrate the expertise from both research fields in a common research endeavor?} We think that activities such as this workshop or our own experience with the ETVIS workshop\footnote{https://www.etvis-workshop.org} and joint research centers (like SFB-TRR~161\footnote{https://www.sfbtrr161.de/}) are a good way to go, but are alone not sufficient and need further action. Building a research area of visualization psychology could be a viable means, for example, by establishing publication and other presentation opportunities that work for visualization researchers, psychologists and social scientists alike, by setting up a canon of teaching new students, and by lobbying for funding possibilities for such interdisciplinary work. 

\paragraph{Psychology Education}

Although many design principles are based on perceptual and cognitive theories, in-depth psychological background knowledge is often not part of the education for visualization. 
Researchers starting with eye tracking studies are confronted with learning eye tracking methodology, which is, starting with proper calibration to a comprehensive analysis of the data, a complex field on its own. As a consequence, deeper knowledge of a whole new research field, i.e., psychology, is hard to achieve within the short time span of an average PhD student's career.

\section{Acknowledgments}

This work was funded by the Deutsche Forschungsgemeinschaft (DFG, German Research Foundation) – project ID 251654672 – TRR 161 (project B01) and under Germany’s Excellence Strategy –
EXC 2120/1 – 390831618.

%
%
\bibliographystyle{abbrv}
\bibliography{template}

\end{document}